\documentclass[prx,aps,twocolumn,showpacs,raggedbottom,eprint,superscriptaddress,10pt,longbibliography,floatfix]{revtex4-2} 
\usepackage[utf8]{inputenc}
\usepackage{braket}
\usepackage{graphicx}
\usepackage{amsmath}
\usepackage{amssymb}
\usepackage{hyperref}
\usepackage{float}
\usepackage{booktabs}
\usepackage{tabularx}
\hypersetup{
	colorlinks=true,
	linkcolor=blue,
	filecolor=blue,
	citecolor = black,      
	urlcolor=cyan,
}

\begin{document}

\title{Experimentally accessible scheme for a fractional Chern insulator in Rydberg atoms}
\author{S.~Weber}
\affiliation{Institute for Theoretical Physics III and Center for Integrated Quantum Science and Technology, Universität Stuttgart, Pfaffenwaldring 57, 70569 Stuttgart, Germany}
\author{R.~Bai}
\affiliation{Institute for Theoretical Physics III and Center for Integrated Quantum Science and Technology, Universität Stuttgart, Pfaffenwaldring 57, 70569 Stuttgart, Germany}
\author{N.~Makki}
\affiliation{Institute for Theoretical Physics III and Center for Integrated Quantum Science and Technology, Universität Stuttgart, Pfaffenwaldring 57, 70569 Stuttgart, Germany}
\author{J.~Mögerle}
\affiliation{Institute for Theoretical Physics III and Center for Integrated Quantum Science and Technology, Universität Stuttgart, Pfaffenwaldring 57, 70569 Stuttgart, Germany}
\author{T.~Lahaye}
\affiliation{Universit\'e Paris-Saclay, Institut d’Optique Graduate School, CNRS, Laboratoire Charles Fabry, 91127 Palaiseau Cedex, France}
\author{A.~Browaeys}
\affiliation{Universit\'e Paris-Saclay, Institut d’Optique Graduate School, CNRS, Laboratoire Charles Fabry, 91127 Palaiseau Cedex, France}
\author{M.~Daghofer}
\affiliation{Institute for Functional Matter and Quantum Technology, and Center for Integrated Quantum Science and Technology, University of Stuttgart, 70569 Stuttgart, Germany}
\author{N.~Lang}
\affiliation{Institute for Theoretical Physics III and Center for Integrated Quantum Science and Technology, Universität Stuttgart, Pfaffenwaldring 57, 70569 Stuttgart, Germany}
\author{H.~P.~B\"uchler}
\affiliation{Institute for Theoretical Physics III and Center for Integrated Quantum Science and Technology, Universität Stuttgart, Pfaffenwaldring 57, 70569 Stuttgart, Germany}

\begin{abstract} 
We present a setup with Rydberg atoms for the realization of a bosonic fractional Chern insulator in artificial matter. The suggested setup relies on Rydberg atoms arranged in a honeycomb lattice, where excitations hop through the lattice by dipolar exchange interactions, and can be interpreted as hard-core bosons.
The quantum many-body Hamiltonian is studied within exact diagonalization and DMRG.  We identify experimentally accessible parameters where all signatures indicate the appearance of a fractional state with the same topological properties as the $\nu=1/2$ bosonic Laughlin  state. We demonstrate an adiabatic ramping procedure, which allows for the preparation of the topological state in a finite system, and demonstrate an experimentally accessible smoking gun signature for the fractional excitations.
\end{abstract}
\maketitle

%%%%%%%%%%%%%%%%%%%%%%%%%%%%%%%%%%%%%%%%%%%%%%%%%%%%%%%%%%%%%%%%%%%%%%%%%%%%%%%%%%%%%%%%%%%%%%

\section{Introduction}

Many-body ground states that feature intrinsic topological order are distinguished by remarkable properties  such as excitations with anyonic statistics, long-range entanglement, and/or robust edge states \cite{Wen2017}. These properties make such topological phases of  high interest for applications in quantum information, e.g., topological quantum computation and topological quantum memory \cite{Nayak2008}. The most prominent topological phases are fractional quantum Hall states which naturally appear for electrons confined in two dimensions in strong magnetic fields \cite{stormer1999}. However, in recent years a focus is on realizing such topological phases in artificial matter \cite{Duan2003,Buchler2005,Baranov2005,Sorensen2005,Micheli2006,Cooper2013,Kraus2013,Yao2013,gerster2017fractional,Verresen2021}. Such a quantum simulator allows to probe their properties with novel tools, and makes them amenable for applications \cite{Altman2021}. Especially platforms based on Rydberg atoms have emerged as highly promising, and first experiments demonstrate a symmetry protected topological phase \cite{leseleuc2019spt}, and signatures of a spin liquid \cite{semeghini2021spinliquid}. Here, we investigate and predict the appearance of a bosonic fractional Chern insulator in an experimentally accessible system with Rydberg atoms.

\begin{figure}[t]
	\includegraphics[width=\linewidth]{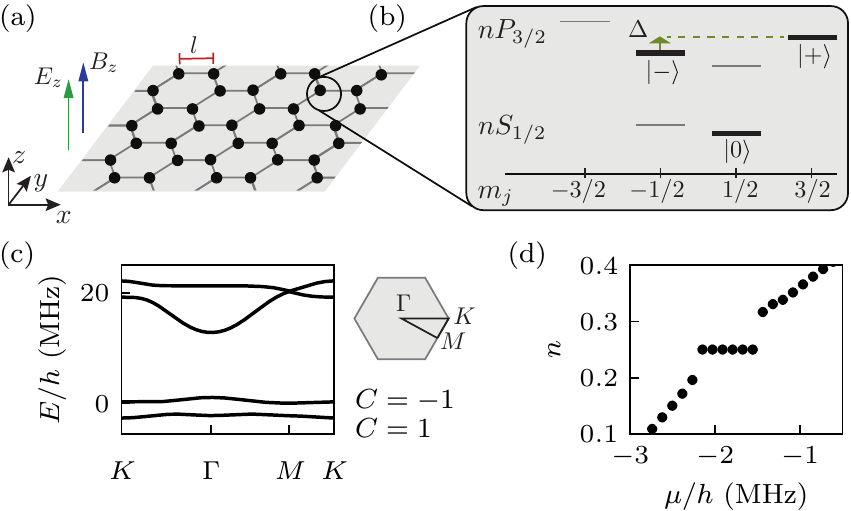}
	\caption{\textbf{Setup for the realization of a fractional Chern insulator.} 
	(a)~Rydberg atoms are arranged in a honeycomb lattice with lattice spacing $l$. A homogeneous electric field $E_z$ and magnetic field $B_z$ are applied perpendicular to the plane of atoms, along the quantization axis $z$.
	(b)~The fields isolate the Rydberg levels of the V-structure $\ket{0}$, $\ket{+}$, and $\ket{-}$ (black lines). The state $\ket{0}$ is treated as the vacuum state and the excitations $\ket{\pm}$ as particles. The energy difference $\Delta = E_{+} - E_{-}$ between $\ket{+}$ and $\ket{-}$ is controlled by the fields.
	(c)~Single-particle band structure along the depicted path through the Brillouin zone for the experimental parameters that are given in the main text ($h$ is Planck's constant). The lowest band has the single-particle Chern number $C = 1$.
	(d)~Average particle density $n$ of the many-body ground state as a function of the chemical potential $\mu$. The density shows a plateau at $1/4$-filling, indicating an incompressible phase.}
	\label{fig:setup}
\end{figure}
    
Rydberg platforms are based on individual atoms trapped by optical tweezers in a programmable array with up to several hundred sites and near perfect unit-filling \cite{Miroshnychenko2006,Barredo2016,Endres2016,Kim2016,Mello2019}. Strong interactions are achieved by exciting the atoms into Rydberg states, and the resulting van der Waals or dipolar exchange interactions enable the realization of different quantum many-body systems. Examples include Ising-like spin models \cite{labuhn2016,Bernien2017,ebadi2021quantum,scholl2021quantum}, and quantum many-body systems based on hard-core bosons  \cite{leseleuc2019spt,lienhard2020realization}. Recently, the appearance of a topological spin liquid in such a Rydberg platform has been predicted \cite{Verresen2021}, and even first experimental signatures of this topological phase have been reported \cite{semeghini2021spinliquid}.
However, the quantum simulation of fractional quantum Hall-like states remains an open challenge. A promising approach is based on combining two main ingredients: strong interactions between particles and the realization of band structures characterized by a Chern number within flat bands and/or homogeneous Berry curvature \cite{Bergholtz2013,jackson2015geometric}. A setup with Rydberg atoms naturally gives rise to effective bosonic particles with a hard-core constraint \cite{leseleuc2019spt}, and the appearance of topological band structures has been predicted for Rydberg excitations hopping under dipolar exchange interaction  \cite{weber2018topologically}.
The key ingredient is the intrinsic spin-orbit coupling of the dipolar exchange interaction, which has been experimentally demonstrated in a setup with three atoms \cite{lienhard2020realization}. 

Here, we present a detailed proposal for the realization of a bosonic fractional Chern insulator with Rydberg atoms that features the same topological properties as the $\nu=1/2$ Laughlin state \cite{Laughlin1983}. We consider a system similar to the models whose single-particle sector gives rise to topological bands with Chern number $C=1$ \cite{weber2018topologically},
and study the ground state properties in the quantum many-body regime within exact diagonalization and DMRG. 
Based on a microscopic analysis, the interactions between the Rydberg states are derived, and we perform an extensive parameter scan to identify an experimentally accessible regime, where the topological phase appears at particle density $n=1/4$.
The key signatures of the topological phase are an excitation gap above two nearly degenerate ground states on a torus, a many-body Chern number $C=1$, as well as a finite topological entanglement entropy. We present an adiabatic ramping scheme, which allows the preparation of the topological state in a finite system with a large excitation gap during the full ramping procedure. Finally, we demonstrate a clear smoking gun signature of the topological
phase, which is accessible with current experimental techniques.

\section{System}

Our system consists of $^{87}$Rb atoms arranged in a two-dimensional honeycomb lattice. For each atom, we consider the Rydberg states $\ket{0}=\ket{nS_{1/2}, m_j = 1/2}$, $\ket{+}=\ket{nP_{3/2}, m_j = 3/2}$, and $\ket{-}=\ket{nP_{3/2}, m_j = -1/2}$. These states form a V-level structure. We apply a homogeneous electric field $E_z$ and magnetic field $B_z$ perpendicular to the plane of atoms, along the quantization axis $z$, and use the resulting Stark and Zeeman shifts to energetically isolate these states from other Rydberg states \cite{weber2018topologically}, see Fig.~\ref{fig:setup}(a,b). The fields also allow for tuning the energy difference $\Delta = E_{+} - E_{-}$ between $\ket{+}$ and $\ket{-}$. We interpret $\ket{0}$ as the vacuum state and an excitation into one of the two other states as a bosonic particle, where $\ket{+}$ and $\ket{-}$ correspond to the possible internal states of the particle. We introduce the bosonic operators $a_i^\dagger$ and $b_i^\dagger$ that create an excitation at lattice site $i$ in the internal state $a_i^\dagger \ket{0} = \ket{+}_i$ and $b_i^\dagger \ket{0} = \ket{-}_i$, respectively. As there is exactly one atom on each lattice site, double occupations are prevented and each lattice site can either be empty or occupied by one particle.
The bosonic operators thus satisfy the hard-core constraint $(a^{\dag}_{i})^{2}=0$ and $(b^{\dag}_{i})^{2}=0$, as well as the mutual constraint $a^{\dag}_{i}b^{\dag}_{i}=0$.

The dipolar exchange interaction between the Rydberg states gives rise to the hopping Hamiltonian \cite{weber2018topologically, peter2015topological}

\begin{align}
H_0 = &\sum_{i \neq j} 
\begin{pmatrix}
a_{i} \\
b_{i} \\
\end{pmatrix}^\dagger
\begin{pmatrix}
-t^a_{ij} & \omega_{ij}\\
\omega_{ij}^{*} & -t^b_{ij} \ \\
\end{pmatrix}
\begin{pmatrix}
a_{j} \\
b_{j} \\
\end{pmatrix}
+\Delta \sum_i n^a_{i}\;, \label{eqn:hopping}
\end{align}
with $n^a_{i}=a_{i}^\dagger a_{i}$. Here, $t^a_{ij}$ and $t^b_{ij}$ are the amplitude of the hopping of a $\ket{+}$-particle and a $\ket{-}$-particle, respectively, between sites $i$ and $j$. The internal state of the particle is conserved by these hoppings. By contrast, the hopping that is associated with the amplitude $\omega_{ij}= |\omega_{i j}| e^{-2 i \phi_{i j}}$ changes a $\ket{+}$-particle into a $\ket{-}$-particle and vice versa. This change of the internal state of the particle is accompanied by the collection of a phase $\pm 2\phi_{ij}$ (spin-orbit coupling), where the angle $\phi_{ij}$ is the polar angle of the distance vector $\mathbf{r}_{ij} = \mathbf{r}_j - \mathbf{r}_i$ between sites $i$ and $j$ \cite{lienhard2020realization}. While this angle depends on the chosen coordinate system, the physically meaningful phase, that is collected on a closed path, is independent of this choice. 

\begin{figure*}[t]
	\includegraphics[width=\textwidth]{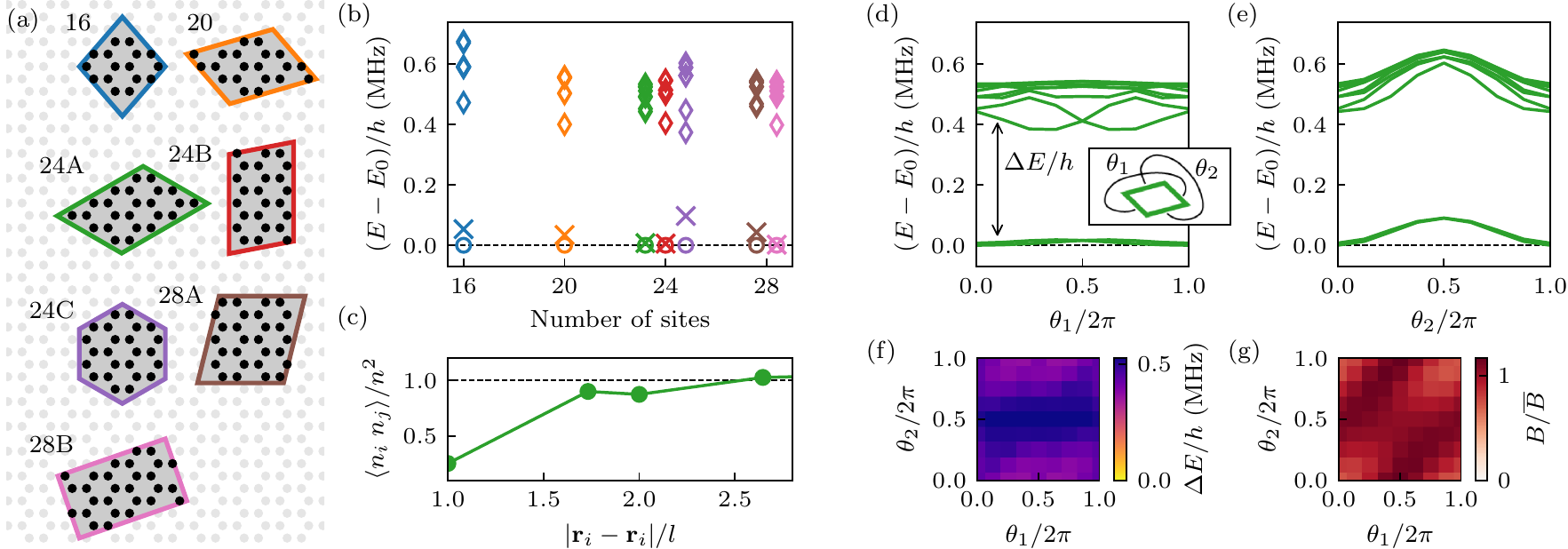}
	\caption{\textbf{Ground state on a torus.} 
	(a)~For $1/4$-filling, the 10 lowest eigenstates are calculated via exact diagonalization for different clusters of the honeycomb lattice with periodic boundary conditions.
	(b)~For all clusters, the ground state is nearly two-fold degenerate and separated from the first excited state by a gap $\Delta E / h \gtrsim 0.2\;\text{MHz}$. $E_0$ is the respective energy of the lowest eigenstate.
	(c)~For the ground state, density-density correlations show no order: The correlation function $\braket{n_i n_j} / n^2$  with density operator $n_i = n^a_{i}+n^b_{i}$ and average density $n$ is short-ranged, and  is depicted exemplarily for the clusters \textit{24A}.
	(d,e)~We now apply twisted boundary conditions with twist angles $\theta_1$ and $\theta_2$ as illustrated in the inset. We plot the lowest eigenenergies as a function of one twist angle while keeping the other zero for the clusters \textit{24A}. The ground state remains quasi-degenerate.
	(f)~The gap to the first excited state stays wide open, independent of the twist angles.
	(g)~The normalized Berry curvature $B/\overline{B}$ of the quasi-degenerate ground state is mostly homogeneous as a function of the twist angles. Here, $\overline{B}$ is the average of the Berry curvature $B$ over all angles. The many-body Chern number is $C = 1$.
	}
	\label{fig:ed}
\end{figure*}

Van der Waals interactions and other higher order interaction processes can give rise to density-density interaction. Another contribution comes from the applied electric field that induces static dipole moments to the Rydberg states. The density-density interactions can be written as 
\begin{align}
H_\text{int} = \frac{1}{2} \sum_{\substack{i \neq j \\ \alpha, \beta \in \{0,a,b\}}} V^{\alpha \beta}_{ij}\;n^{\alpha}_{i} n^{\beta}_{j}\;, \label{eqn:interaction}
\end{align}
with $n^a_{i}=a_{i}^\dagger a_{i}$, $n^b_{i}=b_{i}^\dagger b_{i}$,  $n^0_{i}=1-n^a_{i}-n^b_{i}$, and $V^{\alpha \beta}_{ij}$  the strength of the density-density interaction between sites $i$ and $j$. While in principle additional two-body terms are possible, these terms are two orders of magnitude smaller than the relevant energy scales for the  realistic experimental parameters studied in this manuscript; see Appendix~\hyperref[app:parameters]{B} for an example of such a term. Thus, the full microscopic Hamiltonian reads $H = H_0 + H_\text{int}$.

For the realization of a fractional Chern insulator we find a suitable set of realistic experimental parameters, see Appendix \hyperref[app:realization]{A}. We propose to use the principal quantum number $n=60$, the lattice spacing $l=12\;\mu\text{m}$, the electric field $E_z = 0.725\;\text{V/cm}$, and the magnetic field $B_z = -8\;\text{G}$. For these parameters, we apply the software {\it pairinteraction} \cite{weber2017calculation} to calculate the Stark and Zeeman shifted Rydberg states. Within the basis of these states, we calculate the hopping amplitudes and interactions, and derive the effective Hamiltonian for the relevant levels, see Appendix \hyperref[app:parameters]{B} for details.  %\cite{cohen1998roc} 
For the chosen experimental parameters, the energy difference is $\Delta / h =18.52\;\text{MHz}$ with Planck's constant $h$ and the nearest neighbor hoppings are $t^a / h=1.26\;\text{MHz}$, $t^b / h =0.49\;\text{MHz}$, and $\omega / h =2.38\;\text{MHz}$. For large distances, the hopping amplitudes decrease as $1/|\mathbf{r}_{ij}|^3$ in good approximation. The precise values are given in Appendix \hyperref[app:parameters]{B}, where the values of the density-density interactions are also listed. In general, $| V^{\alpha \beta}_{ij} | \leq 0.3\;\text{MHz}$ and we checked that the results do not change qualitatively if we switch off the density-density interactions. However, to be as close as possible to a potential experimental realization, we keep these terms in our calculations. Throughout the paper, we take into account hoppings and interactions up to next-next-nearest neighbors as longer ranging processes would cause issues due to self-interaction in systems with periodic boundary conditions as studied later on.

For now, let us focus on the single-particle band structure. Due to the two-site unit cell of the honeycomb lattice and the two possible internal states of a particle, it has four bands. The external magnetic field breaks  time-reversal symmetry, enabling topologically non-trivial bands. Indeed, the lowest band has a non-zero single particle Chern number \cite{fukui2005chern} $C = 1$, see Fig.~\ref{fig:setup}(c). The fluctuations of the Berry curvature over the Brillouin zone, which are quantified by their root-mean-square value $\sigma_B=0.4$, are small; see \cite{jackson2015geometric} for the precise definition of $\sigma_{B}$. It has been found that similar honeycomb systems can feature rather flat bands \cite{peter2015topological}. This is also the case for the experimental parameters we propose, where the lowest band has a flatness ratio $f=2.7$; here, $f$ is the ratio of band gap divided by band width. In combination with the strong on-site interaction due to the hard-core constraint, these properties of the single-particle band structure make our system a promising candidate for the realization of a fractional Chern insulator in the many-body regime.

We start our exploration of the many-body regime by applying a chemical potential $\mu$ to the system and calculating the resulting average particle density $n=n^{a}+n^{b}$ of the ground state with density matrix renormalization group (DMRG), using the infinite-DMRG implementation of the open-source software {\it TeNPy} \cite{tenpy}. This method is based on the matrix product state (MPS) approach and allows for the study of our system on an infinite cylinder, see below for more details. The density shows a plateau at $1/4$-filling, indicating an incompressible phase, see Fig.~\ref{fig:setup}(d). 
Note that the filling $n=1/4$ corresponds to $1/2$-filling of the lowest band with Chern number $C=1$, and therefore is compatible with a topological phase exhibiting the same properties as the $\nu=1/2$ bosonic Laughlin state.

%%%%%%%%%%%%%%%%%%%%%%%%%%%%%%%%%%%%%%%%%%%%%%%%%%%%%%%%%%%%%%%%%%%%%%%%%%%%%%%%%%%%%%%%%%%%%%

\section{Topological order}

In the following, we study the many-body ground state at $1/4$-filling and demonstrate that it indeed shows the characteristic properties of a bosonic fractional Chern insulator.

\subsection{Exact diagonalization on a torus}

We start by analyzing the ground state properties with exact diagonalization. For this analysis, we consider various clusters \cite{varney2012quantum} of the honeycomb lattice with periodic boundary conditions, see Fig.~\ref{fig:ed}(a). The chosen clusters are different tessellations of the honeycomb lattice, and consist of $L=16$ to $L=28$ sites. Periodic boundary conditions impose the  topology of a torus on a cluster.
Thus, for our model, where the particles have two internal states, the resulting many-body bases comprise $2^N L!/N!(L-N)!$ states, which for $L=28$ lattice sites corresponds to $\sim 152$ million states at $1/4$-filling with $N=7$ particles. Using exact diagonalization, we calculate the 10 lowest eigenstates and find that the ground state of our system fulfills three characteristic features of a bosonic fractional Chern insulator:

\begin{itemize}
\item First, the ground state is nearly two-fold degenerate and separated from the first excited state by a gap $\Delta E / h \gtrsim 0.2\;\text{MHz}$ for all the studied clusters, see Fig.~\ref{fig:ed}(b). These properties are robust under twisted boundary conditions on the torus with twist angles $\theta_1$ and $\theta_2$, see Fig.~\ref{fig:ed}(d-f).

\item Second, all local correlation functions in the ground state decay exponentially in the bulk. As an example, the short ranged behavior of the density-density correlation is shown in Fig.~\ref{fig:ed}(c). Therefore, we find absence of any spontaneous symmetry breaking. 

\item Third, we determine the many-body Chern number for the nearly degenerate two-fold ground state manifold. The approach is  based on applying twisted boundary conditions and determining the corresponding Berry curvature \cite{hatsugai2005characterization}. The Berry curvature is mostly homogeneous, see Fig~\ref{fig:ed}(g), and we find the many-body Chern number $C=1$.
\end{itemize}

These three observations are a clear indication of a ground state exhibiting topological order with long-range entanglement. Furthermore, all these observations are the characteristic topological properties of a $\nu=1/2$ bosonic Laughlin state \cite{Laughlin1983,wang2011fractional,gerster2017fractional,Rosson2019}, which is the simplest topological phase for bosons in a half-filled topological band and has been predicted in closely related systems \cite{wang2011fractional}.  In order to demonstrate the long-range entanglement, we determine the topological entanglement entropy with DMRG in the next section.

\begin{figure}[htb]
	\includegraphics[width=\linewidth]{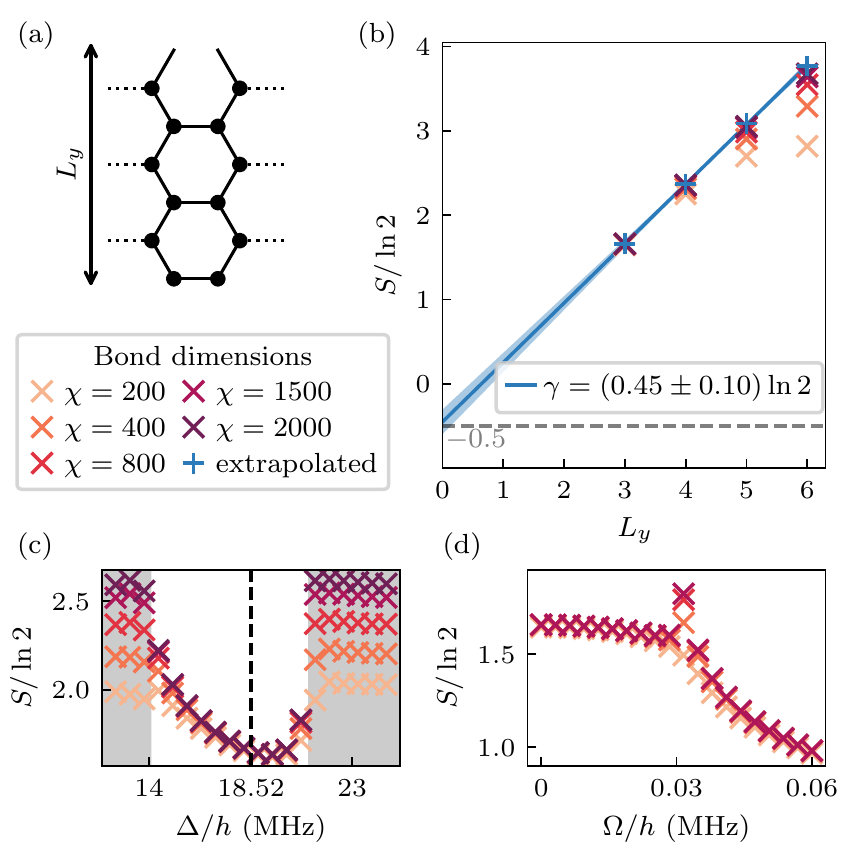}
	\caption{\textbf{Scaling of the entanglement entropy.}
	(a) The unit cell used for infinite-DMRG calculations. We vary $L_y$ from $3$ up to $6$ (i.e. a maximum of $24$ sites) and apply periodic boundary conditions in $y$ direction. In $x$ direction the unit cell repeats forming an infinite cylinder.
	(b) Area law behavior of the entanglement entropy for an equal bipartition of the infinite cylinder and different cylinder circumferences.
	Bond dimensions $\chi$ up to $2000$ are used to extrapolate the entropy for $\chi \rightarrow \infty$, (see Appendix \hyperref[app:entanglemententropy]{C}), from which then the topological entanglement entropy  $\gamma = (0.45 \pm 0.1) \ln 2$ is determined. The error is estimated by calculating $\gamma$ for the entropies $S(\chi = 2000)$ and corresponds to the shaded blue region.
	(c) Entanglement entropy for varying energy difference $\Delta$. The phase transition from the topological phase into a gapless phase (shaded grey) is signaled by the divergence of the entanglement entropy for increasing bond dimension. The dashed line indicates the proposed experimental value $\Delta/h=18.52~\text{MHz}$ for the realistic setup.
	(d) Entanglement entropy for varying Rabi frequency $\Omega$ as required for adiabatic preparation. The gap closing is identified by a diverging peek in the entanglement entropy and determines the critical value $\Omega_{c}/h= 0.032 {\rm MHz}$. As the particle number is no longer fixed, we set the chemical potential  $\mu/h = -1.84~\text{MHz}$.
	}
	\label{fig:dmrg}
\end{figure}

\subsection{Infinite-DMRG}

We analyze the ground state properties within infinite-DMRG, using {\it TeNPy} \cite{tenpy}. The chosen unit cell on the honeycomb lattice for the MPS is shown in Fig~\ref{fig:dmrg}(a) with the width $L_y$ of the cylinder varying from $3$ to $6$, i.e., up to 24 lattice sites, and convergence is checked by studying bond dimensions up to $\chi=2000$. A main result has already been pointed out above and is shown in Fig.~\ref{fig:setup}(d): the particle density $n=n^{a}+n^{b}$ for varying chemical potential exhibits a clear plateau at quarter filling $n=1/4$, and demonstrates the existence of an incompressible phase, which we identify as a bosonic fractional Chern insulator. From the width of the incompressible plateau, we can identify a single particle excitation gap $\sim 0.3\;\text{MHz}$, which is in the same range as the excitation gap for a fixed number of particles found in exact diagonalization. 

Next, we focus on a detailed analysis of the ground state with infinite-DMRG for fixed particle density $n=1/4$.
Especially, we study the entanglement entropy $S$ between the left and the right half of an infinite cylinder by cutting the cylinder into two halves. We find a fast convergence of the entanglement entropy $S$ with increasing bond dimension $\chi$ confirming the presence of an excitation gap. For such a gapped phase, it is well established that the entanglement entropy follows an area law behavior with the scaling behavior for large $L_y$ \cite{Kitaev2006,Levin2006}
\begin{equation}
    S = a  L_y - \gamma\;. 
\end{equation}
Here, $\gamma$ describes the topological entanglement entropy, which  becomes finite in a topological phase with long-range entanglement. We extract this topological term $\gamma$ by studying the behavior of the entanglement entropy for increasing transverse width $L_y$ of the cylinder, see Fig.~\ref{fig:dmrg}(b). As the convergence of the entanglement entropy becomes slower for increasing width $L_y$, we apply a fitting procedure to estimate the entanglement entropy, see Appendix \hyperref[app:entanglemententropy]{C}. We find a non-vanishing topological entanglement entropy $\gamma = (0.45 \pm 0.1) \ln 2$, which is consistent with the  value $\gamma=1/2\: \ln 2$ for the Laughling state at filling $\nu=1/2$ \cite{Kitaev2006,Levin2006,gerster2017fractional,Rosson2019}. 

Finally, we study the stability of this topological phase. The most natural tuning parameter is the energy difference $\Delta$ between the $\ket{+}$ and $\ket{-}$ excitation, which is controlled by the strength of the electric field. In Fig.~\ref{fig:dmrg}(c), we show the entanglement entropy for $12~\text{MHz} < \Delta / h < 25~\text{MHz}$; the latter  corresponds to a variation in the electric field $0.67~\text{V/cm} < E_z < 0.79~\text{V/cm}$, see Appendix \hyperref[app:realization]{A}. We find a clear phase transition from the gapped topological phase,  where the entanglement entropy  exhibits an area law and converges quickly to a finite value determined by the circumference of the cylinder, to a gapless phase with a slowly diverging entanglement entropy for increasing bond dimension; the most prominent candidate for this gapless phase is a superfluid. Remarkably, the topological phase is stable over a wide range of energy differences $\Delta$, which demonstrates robustness under realistic experimental imperfections.

%%%%%%%%%%%%%%%%%%%%%%%%%%%%%%%%%%%%%%%%%%%%%%%%%%%%%%%%%%%%%%%%%%%%%%%%%%%%%%%%%%%%%%%%%%%%%%

\section{Experimental preparation}

\begin{figure}[t]
	\includegraphics[width=\linewidth]{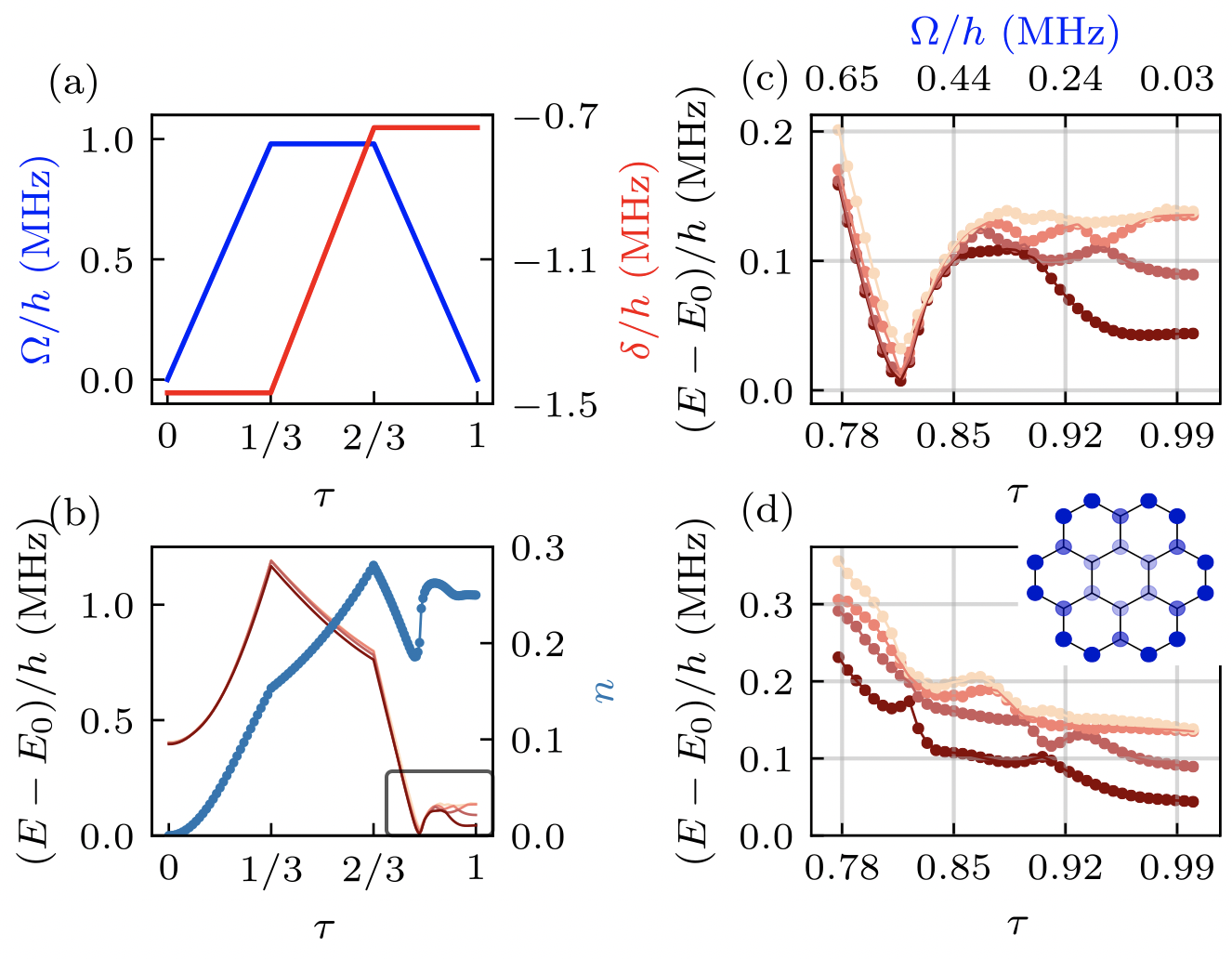}
	\caption{\textbf{Adiabatic preparation of the ground state.} 
(a) Parameter sweep profile showing the three regimes of varying the Rabi frequency $\Omega$ and the detuning $\delta$ as a function of the sweep parameter $\tau$. 
(b) For the 24 site lattice shown in the inset of (d), we plot the energy gaps between the lowest four excited states and the ground state, as well as the average particle density $n$ for the duration of the sweep. At $\tau=0$, the system has a wide excitation gap and a particle density $n=0$. The transition into the topological phase occurs during the final stage of the parameter sweep, where $\Omega$ decreases and $\delta/h$ is fixed at $-0.73$ (MHz), as indicated by the gap closing around $\tau \approx 0.8$. At the end of the parameter sweep, the ground state is quarter filled ($n(1)=0.25$).
(c) Zooming into the transition region, we determine the critical point of the effective model which occurs around $\Omega/h \approx 0.5$ (MHz), and observe a finite size gap $(E_1-E_0)/h=0.05$ (MHz). 
(d) The gap closing is smoothed out in the case where $\Omega(\tau)$ is spacially inhomogeneous. The intensity in color of each site of the lattice shown in the inset correlates to the value of the Rabi frequency this site experiences. The intensity is $100\%$ for outermost sites and drops to $75\%$ and $35\%$ as we get closer to the center.
}
	
	\label{fig:preparation}
\end{figure}

The preparation of the fractional Chern insulator in a closed system, such as the experimental realization with 
Rydberg atoms discussed here, requires an adiabatic sweep with a time-dependent Hamiltonian 
from a trivial state into the topological phase. The most natural initial state is the vacuum state with 
all Rydberg atoms in the level $|0\rangle$. Here, we propose to apply a coupling of the state $|0\rangle$ to $|-\rangle$ by a microwave field with time dependent Rabi frequency $\Omega(\tau)$ and detuning $\delta(\tau)$, 
\begin{equation}
  H_{c} = \sum_{i} \Omega(\tau) \Big[ b^{\dag}_i+b_{i} \Big] - \sum_{i} \delta(\tau) \Big[n^{a}_i+n^{b}_i\Big].
\end{equation}
A sketch of the proposed time dependence is shown in Fig.~\ref{fig:preparation}(a) with the system initially  prepared into the trivial state at $\tau=0$, and one ends in the topological state at $\tau=1$. Here, we introduced the relative time $\tau=t/T$ with $T$ the total time of the applied pulse.
The detuning $\delta(1)$ at the end of the pulse corresponds to the chemical potential $\mu$ for the setup. Its value is determined by the condition to end in the topological phase. In the following, we choose $\delta(1)/h= -1.84 \ {\rm MHz}$, see Fig.~\ref{fig:setup}(d). 

For such an adiabatic sweep, we start in a topologically trivial gapped state and remain in a gapped state during the adiabatic sweep except at a critical coupling strength $\Omega_c$, where the gap closes and the phase transition into the topological phase takes place. Note that such a gap closing is unavoidable for the  adiabatic preparation of a topological phase with long-range entanglement. 
We start by deriving  the critical Rabi frequency $\Omega_c$ within infinite-DMRG on a cylinder. The closing of the gap is signaled by a sharp divergence in the entanglement entropy for increasing bond dimension, see Fig.~\ref{fig:dmrg}(d), and we estimate the critical Rabi frequency $\Omega_{c}/h= 0.032 {\rm MHz}$ at fixed detuning $\delta(1)/h =- 1.84 \  {\rm MHz}$. 

In a next step, we analyze the adiabatic sweep for realistic systems with open boundary conditions and analyze the finite size gap at the transition; the latter limits the speed for the adiabatic sweep. For this purpose, we perform again exact diagonalization, i.e., at each fixed time $\tau$ we determine the ground state and the gap to the first few excited states. As the number of excitations is no longer conserved, the system sizes accessible within exact diagonalization are limited to 16 sites.  To study larger systems, we resort to an effective model describing our system by adiabatically eliminating the Rydberg $\ket{+}$ state, and keeping only the leading terms. This effective model reduces to \cite{lienhard2020realization}
\begin{align}
\frac{H}{t_b} = &- \sum_{\langle i,j \rangle}  b_i^\dagger b_j  - \zeta \sum_{\langle \langle \langle i,j \rangle \rangle \rangle} \ b_i^\dagger b_j 
+  \lambda \sum_{\langle i,j \rangle}   \ n_i n_j \label{eqn:Heff}\\
&-\sum_{\langle \langle i,j \rangle \rangle}  \left ( \frac{1}{3^{3/2}} + \lambda e^{i \frac{2 \pi}{3} p_{ij}} \left ( 1- n_{\bar{ij}}\right )  \right ) b_i^\dagger b_j + h.c. 
\nonumber
\end{align}
Here, again the operator $b^{\dag}_{i}$ creates a $\ket{-}$-excitation at site $i$, i.e., $b_i^\dagger \ket{0} = \ket{-}_i$, while $t_b/h= 0.49\ {\rm MHz }$ denotes the nearest-neighbor hopping strength. The factor $\frac{1}{3^{3/2}}$ accounts for the reduced strength of next-nearest neighbor hopping;
$p_{ij}=\pm 1$ depends on the direction of the hopping process, i.e., it is positive for a clockwise hopping around a hexagon and negative for anti-clockwise hopping, and gives rise to an induced effective magnetic flux. Finally,  $n_{\bar{ij}}$ is the occupation of the site between $i$ and $j$. 
The parameter $\lambda = w^2/\Delta t_{b} $ accounts for the terms due to adiabatic elimination. For the appearance of the topological phase,
we are in a parameter regime  where the adiabatic elimination is no longer fully justified, and with the parameters above, the effective model
does not show a topological phase. However, it is possible to add a next-next nearest-neighbor hopping with strength $\zeta= -0.207$ and choose $\lambda = 0.254$ such that the single particle band structure of the effective model exhibits again a large flatness. Then, we find the same bosonic fractional Chern insulator within the effective model at $1/4$-filling as for the full model. 
Note that in a closely related model, the appearance of such a fractional Chern insulator state for hard-core bosons has been previously predicted  \cite{wang2011fractional}.

We expect that the effective model in Eq.~(\ref{eqn:Heff}) captures the qualitative features of our full system and allows us to study the adiabatic preparation for much larger systems. As shown in Fig.~\ref{fig:preparation}(a), the proposed parameter sweep profile consists of three segments: First, $\Omega$ increases from zero while $\delta$ is fixed at $\delta(0)$. Second, $\Omega$ is fixed while $\delta$ increases to $\delta(1)$. Finally, $\Omega$ decreases back to zero while $\delta$ is fixed. The initial and final values of the detuning $\delta(0)/h$ and $\delta(1)/h$ are chosen to be $-1.47 \ {\rm MHz}$ and $-0.73 \ {\rm MHz}$, respectively. This choice guarantees that we start in the trivial phase with all atoms prepared in the state $\ket{0}$ and end in the topological phase.  In Fig.~\ref{fig:preparation}(b-c), we show the excitation gaps  performing exact diagonalization for a system with $24$ sites for a geometry as shown in the inset. Again, we find a clear closing of the excitation gap at the transition into the topological phase at a critical Rabi frequency $\Omega_c$ with a residual finite size gap $\Delta E/h \approx 0.01 \ {\rm MHz}$. It is this finite size gap which limits the speed of the adiabatic ramping procedure close to the critical point.
Note that for open boundary conditions the ground state in the topological phase is unique with a bulk gap and edge modes with a finite size gap. 

Remarkably, we find that using an appropriate inhomogeneous Rabi frequency, see Fig.~\ref{fig:preparation}(d), the excitation gap remains large for the full ramping procedure. The choice is to reduce the  Rabi frequency in the center and continuously increase it towards the boundary.
The intuitive interpretation for this phenomena is that  the transition into the topological phase occurs first in the center and then grows towards the boundary. Therefore,  excitations are continuously pushed towards the boundary. We expect therefore that such a ramping procedure with an inhomogeneous Rabi frequency allows for a much more efficient preparation of the topological phase even for larger systems. Experimentally, such an inhomogeneous Rabi frequency can be achieved by the combination of a microwave field with an optical Raman transition; for the latter, the Rabi frequency can be varied over distances comparable to the lattice spacing,  see Appendix~\hyperref[app:rabi]{D}.

%%%%%%%%%%%%%%%%%%%%%%%%%%%%%%%%%%%%%%%%%%%%%%%%%%%%%%%%%%%%%%%%%%%%%%%%%%%%%%%%%%%%%%%%%%%%%%

\section{Experimental detection}
\begin{figure}[t]
	\includegraphics[width=\linewidth]{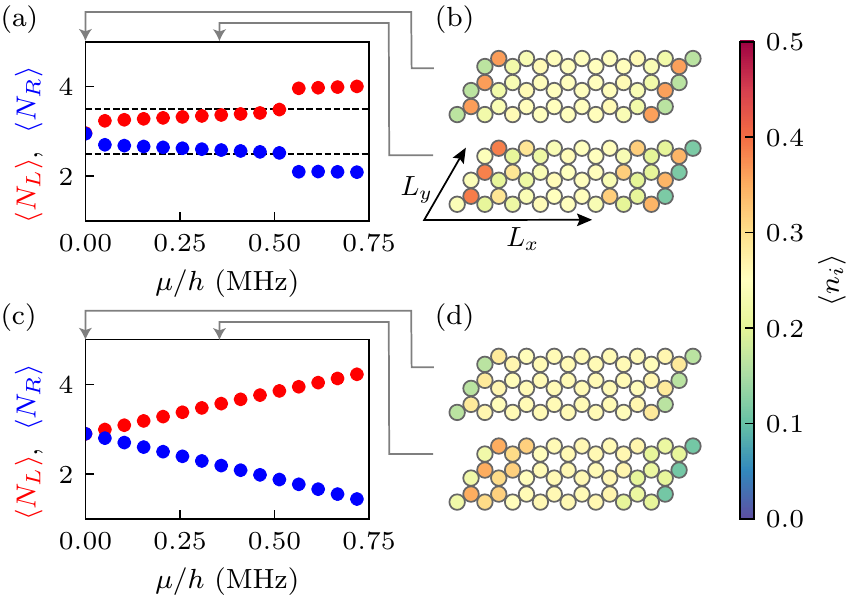}
	\caption{\textbf{Detection of fractional charges.} We consider a cylinder with $L_x = 8$ and $L_y=3$ that is finite along $x$ direction. 
	We apply a chemical potential $\mu \geq 0$ to the 12 leftmost sites $\mathbb{S}_L$ and the opposite potential to the 12 rightmost sites $\mathbb{S}_R$.
	Using DMRG with bond dimension $\chi=200$, we calculate the number of particles on the left $\braket{N_L}$ and on the right $\braket{N_R}$ as a function of $\mu$ with $N_{L,R} = \sum_{i \in \mathbb{S}_{L,R}} n_i$.
	(a)~The number of particles $\braket{N_{L,R}}$ jumps in steps of $1/2$, indicated by the dashed lines.
	(b)~The particle density $\braket{n_i}$ increases on the leftmost sites and drops on the rightmost ones if $\mu$ is increased.
	(c)~For comparison, we study our system in the topologically trivial sector, where the number of particles $\braket{N_{L,R}}$ changes continuously. The system was brought into the trivial sector by increasing $\Delta / h$ by $50~\text{MHz}$ so that the $\ket{+}$ state can be neglected.
	(d)~Corresponding changes in the particle density~$\braket{n_i}$.
	}
	\label{fig:detection}
\end{figure}

While quantities like ground state degeneracy, many-body Chern number, and entanglement entropy can be calculated numerically, they are hard to access in real world experiments. However, recently, an experimentally feasible scheme for the detection of fractionally charged excitations has been proposed \cite{ravciunas2018creating}. We adapt the proposal to our Rydberg setup. For fractional Chern insulators one expects the accumulation of fractionally charged excitations near engineered local defects, and the charge (i.e., the particle number) is easy to access experimentally. 
It is expected to be $1/2$  for a state exhibiting the topological properties of a $\nu=1/2$ Laughlin state \cite{Laughlin1983,Arovas1984}. 

To engineer local defects, we propose to locally apply light shifts to the $\ket{-}$ state which gives rise to a local chemical potential $\mu$. We simulate this scheme using the full microscopic Hamiltonian $H$ described in Eq.~(\ref{eqn:hopping}) and (\ref{eqn:interaction}). While in an experiment one might realize a system with open boundary conditions and several hundreds of atoms \cite{ebadi2021quantum, scholl2021quantum}, we consider a finite cylinder of 48 sites to keep the system accessible with DMRG. The periodic boundary condition along one direction helps us to mitigate finite-size effects. An unrolled version of the cylinder is shown in Fig.~\ref{fig:detection}(b,d). We apply a chemical potential $\mu \geq 0$ to the 12 leftmost sites and the opposite potential to the 12 rightmost sites.

We calculate the ground state at $1/4$-filling as a function of $\mu$ and observe that the particle number summed over the leftmost sites jumps up by $~1/2$ (quasiparticle), while the particle number on the rightmost sites jumps down by the same fraction (quasihole), see Fig.~\ref{fig:detection}(a). The creation of the quasiparticle and quasihole happens already for small values of $\mu$ because the spectrum is gapless at the edge of the cylinder, except for a finite-size gap. The change of the particle number is slightly smaller than $1/2$ because some weight of the quasi-particles also extends into the bulk.

For comparison, we study our system in the topologically trivial sector. To get into the trivial sector, we increase $\Delta / h$ by $50~\text{MHz}$ so that the $\ket{+}$ state can be neglected and we obtain a topologically trivial two-band model. Such a huge increase in $\Delta$ can be experimentally realized, for example, by inverting the direction of the magnetic field $B_z$. We find that in the trivial sector, the number of particles changes continuously, see Fig.~\ref{fig:detection}(c).

%%%%%%%%%%%%%%%%%%%%%%%%%%%%%%%%%%%%%%%%%%%%%%%%%%%%%%%%%%%%%%%%%%%%%%%%%%%%%%%%%%%%%%%%%%%%%%

\section{Conclusion and Outlook}

In this work, we presented a blueprint for the realization of a bosonic fractional Chern insulator with Rydberg atoms. The suggested setup relies on Rydberg atoms arranged in a honeycomb lattice and subject to dipolar exchange interactions, giving rise to hard-core bosons hopping in an effective magnetic field. We performed extensive numerical studies, providing five characteristic signatures for the existence of a state with the same topological properties as a $\nu=1/2$ bosonic Laughlin state
 in our realistic microscopic model at $1/4$-filling: (i) incompressibility of the phase, (ii) nearly two-fold robust ground state degeneracy on a torus, (iii) exponential decay of local correlations in the bulk, (iv) many-body Chern number of $C=1$, and (v) topological entanglement entropy of $\gamma = (0.45 \pm 0.1) \ln 2$. The parameters are chosen for a realistic experimental setup and the analysis is performed for the full microscopic system. 

Remarkably, we find an adiabatic ramping scheme with a spatially inhomogeneous Rabi frequency  where the finite size excitation gap remains large for the full sweep, which allows for a realistic experimental preparation of the ground states.
Finally, we demonstrate a clear smoking gun signature of the topological phase, which is accessible with current experimental techniques. The signature is based on the detection of fractional charges proposed in \cite{ravciunas2018creating}, and numerical simulations provide evidence that this technique can be applied for detecting the Laughlin-like state of  our model.
Our detailed proposal paves the way for the quantum simulation of fractional Chern insulators with Rydberg atoms. Because of the microscopic control of the particles within a quantum simulator, this can help to deepen our understanding of topological states of matter.

\acknowledgments{
This project has received funding from the European Union’s Horizon 2020 research and innovation program under Grant Agreement No. 817482 (PASQuanS), the ERC Advanced grant No 101018511 (ATARAXIA),` as well as the French-German collaboration for joint projects in NLE Sciences funded by the Deutsche Forschungsgemeinschaft (DFG) and the Agence National de la Recherche (ANR, project RYBOTIN). We also gratefully acknowledge financial support by the Baden-W\"urttemberg Stiftung via Grant BWST ISF2019-017 under the program Internationale Spitzenforschung. H. P. B. and A. B. thank the KITP for hospitality. This research was also supported in part by the National Science Foundation under Grant No. NSF PHY-1748958.}

\appendix

\section{Details on experimental parameters}\label{app:realization}

\begin{figure}[b]
	\includegraphics[width=\linewidth]{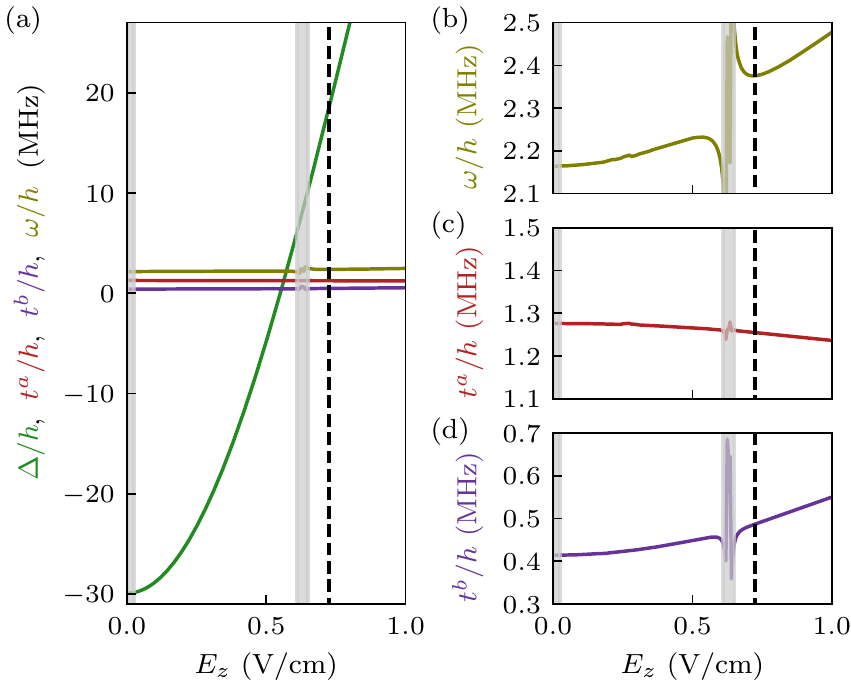}
	\caption{\textbf{Dependence of parameters of the Hamiltonian on the electric field.}
	(a)~The energy difference $\Delta$ depends strongly on the electric field $E_z$ through the Stark effect. In comparison, the nearest-neighbor hopping amplitudes only weakly depends on $E_z$. The dashed lines indicate the value of $E_z=0.725\;\text{V/cm}$ that was used throughout the paper. Note that for some other values, other Rydberg pair states get resonant and are admixed to the states of the V-structure (gray regions indicate an admixture $>5\%$). There, our perturbative calculation of the hopping amplitudes is no longer valid.
	(b-d)~Close-up views of the electric field dependence of the nearest-neighbor hopping amplitudes.
	}
	\label{fig:params}
\end{figure}

In this section, we explain how the parameters of the Hamiltonian $H$ in Eq.~(\ref{eqn:hopping}) and (\ref{eqn:interaction}) depend on the experimental parameters and motivate the chosen parameter set.

\begin{figure}[b]
	\includegraphics[width=\linewidth]{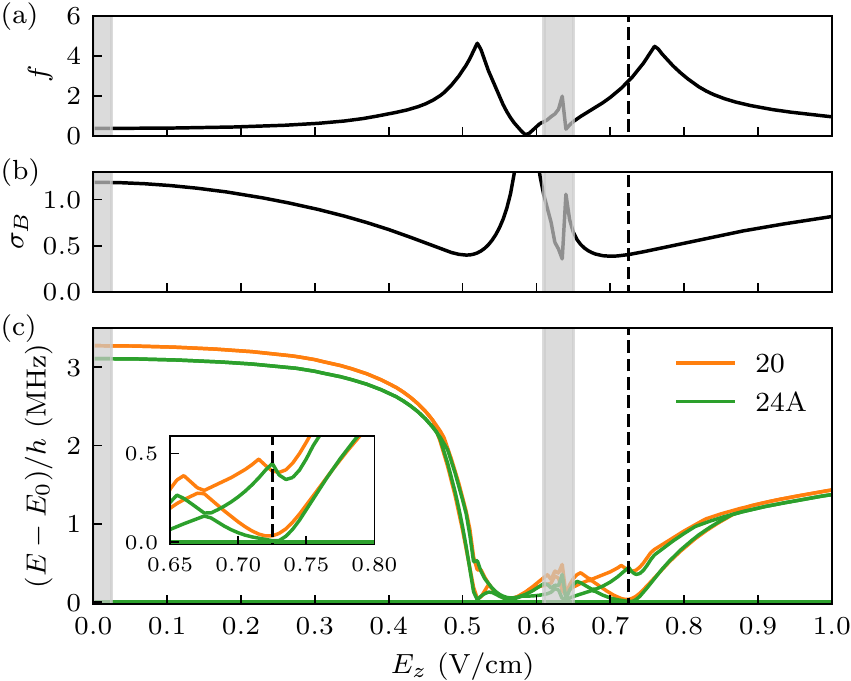}
	\caption{\textbf{Optimization of the electric field.}
	(a)~Flatness ratio $f$ of the lowest band of the single particle band structure as a function of the electric field $E_z$ (gray regions indicate regimes where our perturbative calculation of hopping amplitudes failed). A large flatness ratio is beneficial for realizing a fractional Chern insulator in the many-body regime.
	(b)~Berry curvature fluctuations $\sigma_B$. Typically, a small value is desirable. Note that $f$ and $\sigma_B$ reach their respective optimal values for different $E_z$. Thus, we have to make a compromise.
	(c)~To figure out the optimal value of $E_z$, we calculate the three lowest eigenenergies on a torus at $1/4$-filling for the clusters \textit{20} and \textit{24A}. The ground state is nearly two-fold degenerate in a small region around the optimal value $E_z=0.725\;\text{V/cm}$ (dashed lines). For a zoom into this region, see inset.
	}
	\label{fig:params2}
\end{figure}

We performed our calculations for the principal quantum number $n=60$ as the corresponding Rydberg states have already been studied experimentally in the past. For our choice of the lattice spacing, $l=12\;\mu\text{m}$, we took care that it is large enough to ensure that interactions can be described perturbatively. At the same time, it is small enough to allow for dipolar exchange interactions (order of a MHz) being much faster than the Rydberg decay (the natural lifetime is on the order of $100\;\mu\text{s}$). The magnetic field $B_z = -8\;\text{G}$ was applied to break the Zeeman degeneracy. For the realization of our V-level structure, it is crucial that the sign of the field is negative so that we can compensate for the huge resulting energy splitting between $\ket{+}$ and $\ket{-}$ by applying an electric field of about $E_z = 0.7\;\text{V/cm}$. We can use $E_z$ to fine-tune $\Delta$ without affecting the other parameters much, see Fig.~\ref{fig:params}.

For finding the optimal value for the electric field $E_z$, we first calculate the flatness ratio $f$ and Berry curvature fluctuations $\sigma_B$ as a function of $E_z$, see Fig.~\ref{fig:params2}(a,b). A large $f$ and a small  $\sigma_B$ are beneficial for realizing a fractional Chern insulator in the many-body regime. However, these two quantities reach their respective optimal values for different $E_z$ --- we have to make some compromise. We calculate the three lowest eigenenergies on a torus at $1/4$-filling for the clusters \textit{20} and \textit{24A} to determine the optimal value of $E_z$, see Fig.~\ref{fig:params2}(c). The ground state is nearly two-fold degenerate in a small region around $E_z=0.725\;\text{V/cm}$. Remarkably, this is not the case for the region around $E_z=0.52\;\text{V/cm}$, despite promising values of $f$ and $\sigma_B$. We attribute the lack of a two-fold degenerate ground state for this value of the electric field to the fact that there, the two lowest bands are not energetically separated from the other bands. The determinant condition, which holds for two band models, might potentially be violated \cite{jackson2015geometric}. Thus, $E_z = 0.725\;\text{V/cm}$ is the optimal value for the electric field.

Note that the proposed parameter set is not unique. We can, for example, use a different principal quantum number if we scale the other experimental parameters accordingly.

\section{Numerical calculation of the parameters of the Hamiltonian}\label{app:parameters}

This section contains details on the numerical calculation of the parameters of the Hamiltonian and their values that we used throughout the paper.

We first calculated the effect of the electric and magnetic fields on the Rydberg states. For this, we constructed the Hamiltonian of a single Rydberg atom in the presence of the fields using the {\it pairinteraction} software \cite{weber2017calculation}. We took into account states in the fine structure basis that are at most $80~\text{GHz}$ away in energy from the states of the V-structure $\mathcal{S}=\{\ket{0}, \ket{+}, \ket{-}\}$ and have principal quantum numbers $57 \leq n \leq 63$ and azimuthal quantum numbers $l \leq 4$. By diagonalizing the Hamiltonian, we obtain dressed states that are shifted in energy by the Stark and Zeeman effect. Due to these shifts, the pair states $\mathcal{P}=\{\ket{a,b} \;|\; \ket{a}, \ket{b} \in \mathcal{S} \}$ get energetically isolated from other Rydberg pair states.

Afterwards, we constructed for every pair of atoms the Hamiltonian in the basis of the dressed states. To obtain the effective Hamiltonian that describes the dynamics within the energetically isolated subspace $\mathcal{P}$, we performed a direct rotation that maps the isolated subspace to the space of states that are dressed by the Rydberg-Rydberg interaction \cite{bravyi2011schrieffer}. For this calculation, we ignored Rydberg states that are more than $2~\text{GHz}$ away from the states within $\mathcal{P}$ as their effect on the effective Hamiltonian is insignificant. Because the dressing by the Rydberg-Rydberg interaction is weak (the admixture of other Rydberg states to the states within $\mathcal{P}$ is $<3 \%$), we are in a regime where standard second order perturbation theory is still valid and could have been applied as an alternative approach for calculating the effective Hamiltonian.

Note that in addition to two-body terms, second order processes can lead to three-body terms like correlated hoppings. We checked that these terms are negligible. Additional $n$-body terms might appear in higher order perturbation theory but are in general small except at fine tuned $n$-body resonances.

In the following, we provide the values of the parameters of the Hamiltonian that we have obtained by the aforementioned procedure.  Tab.~\ref{tab:hopping} shows the values of the hopping amplitudes in Eq.~(\ref{eqn:hopping}) and the considered interactions in Eq.~(\ref{eqn:interaction}). All other two-body terms, e.g. $b_{i}^\dagger b_{j}^\dagger a_{i}a_{j}$, are on the order of $0.01\;\text{MHz}$ or less and hence neglected.
\\

\begin{table}[H]
\begin{tabularx}{\linewidth}{XXXX} \toprule
 & $r_{ij}/l=1$ & $r_{ij}/l=\sqrt{3}$ & $r_{ij}/l=2$ \\ \midrule
$t^a_{ij}/h$ & 1.26 & 0.24 & 0.16 \\
$t^b_{ij}/h$ & 0.49 & 0.09 & 0.06 \\
$|\omega_{ij}|/h$ & 2.38 & 0.45 & 0.29 \\
$V^{0,0}_{ij}/h$ & \phantom{-}0.03 & 0.00 & 0.00 \\
$V^{0,a}_{ij}/h$ & \phantom{-}0.07 & 0.01 & 0.01 \\
$V^{0,b}_{ij}/h$ & -0.20 & 0.00 & 0.00 \\
$V^{a,a}_{ij}/h$ & \phantom{-}0.19 & 0.04 & 0.03 \\
$V^{a,b}_{ij}/h$ & \phantom{-}0.25 & 0.05 & 0.03 \\
$V^{b,b}_{ij}/h$ & \phantom{-}0.28 & 0.05 & 0.04 \\ \bottomrule
\end{tabularx}
\caption{Hopping amplitudes and interactions up to next-next-nearest neighbor hopping in MHz. To avoid self-interaction in systems with periodic boundary conditions, we neglect longer ranging processes. Note that $V^{\alpha,\beta}_{ij}=V^{\beta,\alpha}_{ij}$.}
\label{tab:hopping}
\end{table}

\begin{figure}[htb]
	\includegraphics[width=\linewidth]{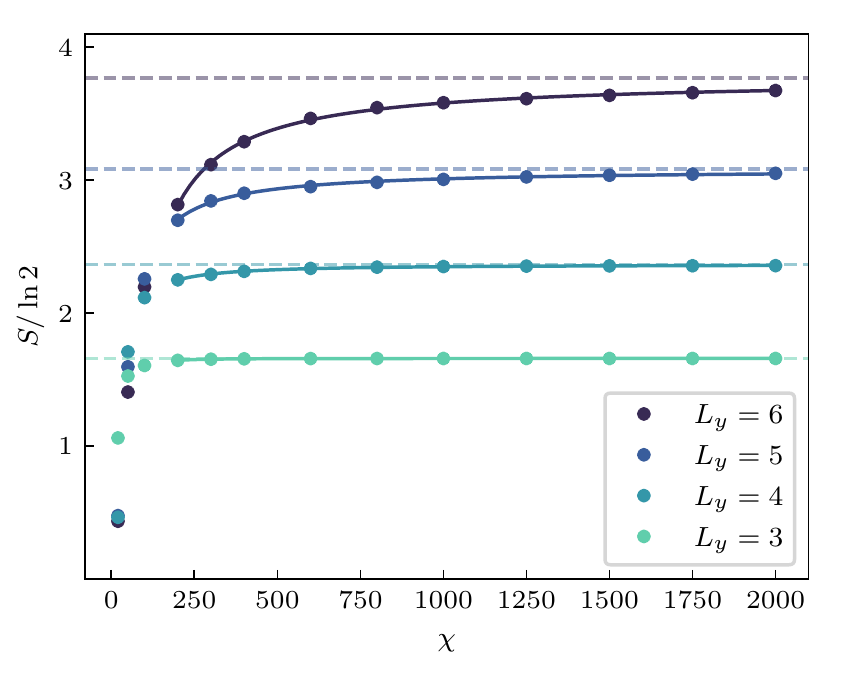}
	\caption{\textbf{Extrapolation of the entanglement entropy to infinite bond dimensions.}
	We used extrapolation functions of the form $S = S_\text{inf} - b \chi^{-c}$ to determine the entanglement entropy $S_\text{inf}$ for $\chi \rightarrow \infty$.
	}
	\label{fig:extrapolation}
\end{figure}
\section{Entanglement entropy extrapolation}\label{app:entanglemententropy}

In this section we discuss the extrapolation of the entanglement entropy to infinite bond dimensions.
In Fig.~\ref{fig:extrapolation} the entanglement entropy is plotted for increasing bond dimensions and all different system sizes $L_y$.
We can clearly see the convergence of the entropy for increasing bond dimensions.
As expected, this convergence gets slower for larger systems, which is why we use a simple fit function $S = S_\text{inf} - b \chi^{-c}$ to extrapolate the entropy $S_\text{inf}$ for $\chi \rightarrow \infty$.
In the fit we ignore all values for bond dimensions $\chi < 200$ to reduce errors from very small bond dimensions. The resulting $S_\text{inf}(L_y)$ are then used to fit an area law $S = a L_y - \gamma$.
From this we then extract the topological entanglement entropy resulting in $\gamma = 0.45 \ln 2$, see Fig.~\ref{fig:dmrg}(b).
We furthermore estimate an error by fitting the area law also to the values $S(\chi=2000)$, which results in $\gamma(\chi = 2000) = 0.35 \ln 2$, yielding an error $\pm 0.1 \ln 2$.
Note that if we only use system sizes $L_y \in \{3, 4, 5\}$, where the entropy converges faster, the resulting values are $\gamma(\chi \rightarrow \infty) = 0.48 \ln 2$ and $\gamma(\chi = 2000) = 0.43 \ln 2$, showing good agreement.

\section{Realization of the spatially inhomogeneous Rabi frequency}\label{app:rabi}

As detailed in the main text, we propose to prepare the ground state via an adiabatic ramping scheme with a spatially inhomogeneous Rabi frequency to keep the finite size gap large. In the following, we outline a possible approach for implementing the spatially inhomogeneous Rabi frequency.

Because spatial inhomogeneities on the micrometer scale cannot be realized by microwave fields, we advise against driving the transition between $\ket{-}$ and $\ket{0}$ directly. Instead, we propose to apply a Raman-like scheme via two intermediate states. We can off-resonantly couple $\ket{-}$ to a lower lying electronic state, for example, $\ket{5D}$ and couple $\ket{5D}$ to a Rydberg state $\ket{n'P}$ using two lasers, while simultaneously driving the transition between $\ket{n'P}$ and $\ket{0}$ using a microwave field. The frequency of the latter can be selected such that $\ket{-}$ and $\ket{0}$ are coupled resonantly by the resulting three-photon transition with an effective Rabi frequency $\Omega_\text{eff}$. By shaping the spatial profile of the laser beam that couples $\ket{5D}$ to $\ket{n'P}$, we can make $\Omega_\text{eff}$ spatially inhomogeneous.

Note that the applied fields lead to AC-Stark shifts of the states $\ket{0}$, $\ket{+}$, and $\ket{-}$. However, because the shifts are nearly homogeneous and can be made the same for $\ket{+}$ and $\ket{-}$ using the possibility to tune the polarization of the laser that couples these states off-resonantly to $\ket{5D}$, the preparation is not affected.

%%%%%%%%%%%%%%%%%%%%%%%%%%%%%%%%%%%%%%%%%%%%%%%%%%%%%%%%%%%%%%%%%%%%%%%%%%%%%%%%%%%%%%%%%%%%%%

\bibliography{manuscript}

%%%%%%%%%%%%%%%%%%%%%%%%%%%%%%%%%%%%%%%%%%%%%%%%%%%%%%%%%%%%%%%%%%%%%%%%%%%%%%%%%%%%%%%%%%%%%%

\end{document}